\documentclass{revtex4}

\usepackage{graphicx}
\usepackage{wrapfig}
\def\Cer{\v{C}erenkov}
\begin{document}
\bibliographystyle{revtex}


\title[K2K Results]{Results from the K2K Long--Baseline Neutrino Oscillation Experiment}



\author{James E. Hill}
\email[jimhill@neutrino.kek.jp]{}
\homepage[]{http://superk.physics.sunysb.edu/~jimhill/}
\affiliation{SUNY Stony Brook, {\rm and} ICRR (University of Tokyo)\\
          \rm for the K2K (KEK-E362) Collaboration}

\collaboration{K2K (KEK-E362) Collaboration}
\noaffiliation

\date{October 15, 2001}

\begin{abstract}
The K2K Long--Baseline neutrino oscillation experiment has been
aquiring data since mid-1999 and has analysed those up to 
March of 2001. Forty-four fully contained events are observedin the fiducial
volume of the far detector where $64^{+6.1}_{-6.6}$
 are expected based partly on measurements
near the beam production point. There is virtually no background
for the contained event search. The methods established in this
experiment are crucial for operation of future similar experiments
to probe the nature of mixing in the neutral lepton sector, a
necessary step in understanding the nature of family structure and of
mass itself.
A brief history and a few notes about the future and direction of
the field precede the description of the experiment and its results.
\end{abstract}

\maketitle

\section{Motivation for long--baseline experimentation} 
\begin{figure}[!ht]
\includegraphics[width=\textwidth]{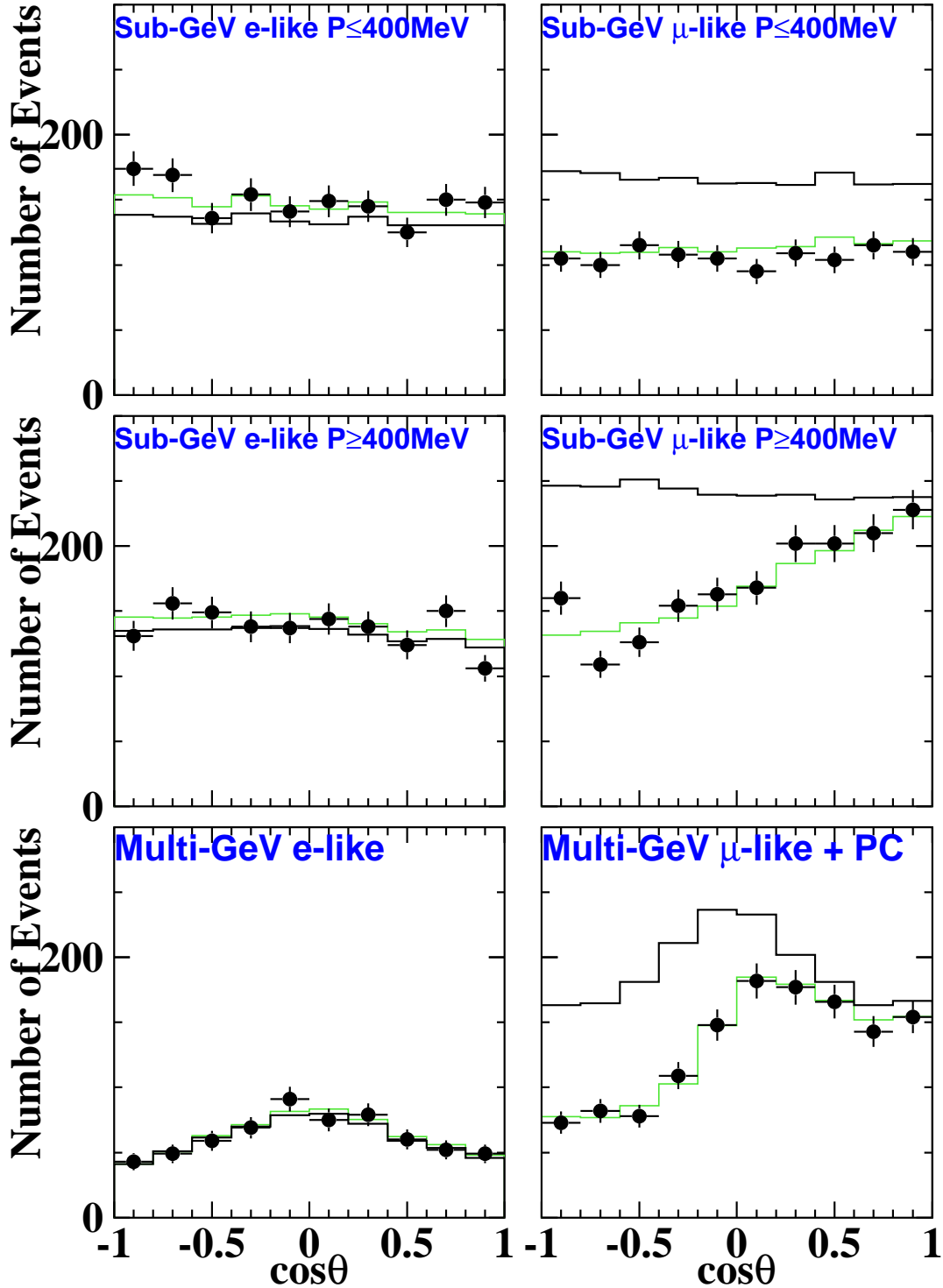}
\caption{The Super--Kamiokande atmospheric neutrino data. The plots are
         described in some detail in the text.}
\label{Plenary_hill_0712_fig1}
\end{figure}

K2K~\cite{Plenary_hill_0712_ref1} is the world's first experiment designed
specifically to search for oscillations in the neutral lepton sector
using the technique of long--baseline experimentation.
 The main motivation for long--baseline accelerator based neutrino
experiments lies in the results provided by the analysis of
atmospheric neutrino data~\cite{Plenary_hill_0712_ref2}.
Figure~\ref{Plenary_hill_0712_fig1} shows the most current information from
Super--Kamiokande regarding the zenith angle distributions of
various event classes in the atmospheric neutrino sample.
A detailed explaination of the plots in the figure is given below.
Earlier atmospheric neutrino experiments generally gave consistent
results although with much lower statistics.

\subsection{Historic  Results Implying Neutrino Oscillation} 
The first observation of atmospheric neutrinos~\cite{Plenary_hill_0712_ref3}
came in the 1960's from underground experiments searching for proton
decay.
Atmospheric neutrino studies began in earnest in the 1980's when
proton decay experimenters realized the need to understand 
their backgrounds better. The first indications of the so-called
``atmospheric neutrino anomaly'' from the water \Cer{} detectors
Kamiokande and IMB~\cite{Plenary_hill_0712_ref4,Plenary_hill_0712_ref5}
were that the observed flavor ratio of atmospheric neutrinos did
not match the expectation.
Originally, the Soudan, NUSEX~\cite{Plenary_hill_0712_ref6}, and
Frejus~\cite{Plenary_hill_0712_ref7} experiments,
which used iron calorimeters, appeared to disagree
with the conclusion from the water target experiments, although
these lacked sufficient statistics to refute any other claim.
Later analyses~\cite{Plenary_hill_0712_ref8}
of of Soudan's full data set data appeared
consistent with the previous water experiments and the
new one, Super--Kamiokande, although still with smaller statistics. 

 The first generation of large water \Cer{} detectors began analyses
of the directional distribution of neutrino events, but lacked the
statistical power to fully exploit the information. When Super--Kamiokande
began taking data in 1996, within a few years it had acquired and analyzed
 a sample of atmospheric neutrino events larger than the sum of all
previous experiments. The larger dimensions of the detector also 
allow us to study higher energy events in which the neutrino reaction
products are all contained within the detector. The higher density of
photo-cathode coverage gives better energy resolution and slightly
better directional reconstruction. Finally, the large volume gives 
us the statistics necessary to study the directional distribution
of neutrino events in sufficient detail to address the issue of
oscillations.

In figure~\ref{Plenary_hill_0712_fig1},
each plot is the azimuthal direction distribution for
a discrete subclass of the atmospheric neutrino event event sample.
Positive values of $\cos(\theta)$ represent
down-going events which have in general come from tens of km away
from the atmosphere above, and negative values represent up-going events
which come from neutrinos created as far as a full earth diameter away. 
In each case the points are data, the thick dark line is the result
of simulations assuming no neutrino oscillation or other new physics,
and the thin light green line is the result of simulations assuming 
oscillations with a parameter set resulting from a fit to these data.
All plots on the left side are for single ring contained events with the 
ring identified as a showering type. Most of these are therefor the
result of charged-current $\rm\nu_e$ or $\overline{\rm\nu}_e$ interactions.
Plots on the right are for single ring contained events with the 
ring identified as a non-showering type, plus (included in the set
in the bottom panel) single ring events with a vertex reconstructed
within the detector fiducial volume, but with the reconstructed track
exiting the inner detector.
In general, the average energy of events represented is highest for events
at the bottom of the page and lowest at the top.
The quantity plotted is the direction cosine for a charged lepton from a 
neutrino interaction, thus for the lowest energy events the direction
is smeared by the wide scattering angle expected in low energy collisions.

While the distributions regarding showers all agree with simulations,
those involving non-showering events do not.
For higher energy events, the distribution is strikingly asymmetric
about the horizon ($\cos(\theta)=0$).
The simplest interpretation is that fewer $\nu_\mu,\overline{\nu}_\mu$
enter the detector from below than from above.
The symmetry of the earth and of the primary cosmic ray distribution
suggest that no such asymmetry can exist for the production of neutrinos
in the atmosphere.
A simple, yet elegant, solution to this would be some process which 
converts neutrinos into a type unobservable in this detector as they
propagate through space.
Oscillations between $\nu_\mu$ and $\nu_\tau$ are an obvious candidate
since the charged current interaction threshold for $\nu_\tau$ is 
relatively high and the small number of interactions that do occur
are unlikely to enter the single ring sample.

\subsection{The Next Steps}
Super--Kamiokande proves rather conclusively that the atmospheric
neutrino results demonstrate some lack of understanding of the underlying
physics. The simplest amendment to the Standard Model of electro-weak
interactions which would explain this is neutrino oscillations.
With this strong evidence and a firm testable hypothesis, the two tasks at
hand for particle physicists are to verify or refute the effect in a controlled
experiment, and to probe the values of relevant parameters of the theory,
possibly with a whole series of experiments each drawing on the 
knowledge and experience gained in the previous.

K2K is designed to explore that first step. Through the 1990's, several
similar experiments were proposed; a few of those proposals are in
this generation of experiments, but only K2K has started data taking.
The experiments which will follow in the next few years,
MINOS~\cite{Plenary_hill_0712_ref9}, the CNGS
projects~\cite{Plenary_hill_0712_ref10}, and some
longer time scale projects generically referred to as
``superbeam'' or ``neutrino factory'' experiments represent a program
of measurements to measure the full neutrino mixing matrix with 
precision competitive with (or in some estimations better than) the
precision of corresponding measurements in the quark sector.

MINOS intends to concentrate on measurement of the parameters of 
$\nu_\mu$ mixing by using a more intense beam than K2K and a more
precisely tuned distance to its far detector. A major goal is to
address the issue of spectral distortions due to oscillations
in a convincing quantitative way allowing a more precise determination
of the mixing parameters.
The CNGS experiments have chosen to concentrate instead on doing
an {\em appearance\/} experiment to look for the $\nu_\tau$ presumably
present in the oscillated beam.
Both of these experiments should start taking data around the year 2005.
They will start from the results of Super--Kamiokande and K2K
and directly build on the technological background of K2K.
In parallel, studies currently underway on low energy neutrinos
from the sun~\cite{Plenary_hill_0712_ref11} and from
reactors~\cite{Plenary_hill_0712_ref12}
will probe another corner of the three generation mixing matrix.
The farther future projects aim to explore the full character of
the $3\times 3$ mixing matrix, including the possibility CP violating
terms just as has been pursued in the quark sector for many years now.

\subsection{The Generic Strategy of Long--Baseline Experimentation} 
The atmospheric neutrino results --our main motivating measurement--
have as their source mostly neutrinos in the GeV momentum range 
which travel geographical distances.
Thus, a controlled experiment to
check its validity might use accelerator produced neutrinos and a 
detector at a well known distance some hundreds of kilometers away.
To ensure full understanding of the beam at production, a near detector
is useful; to reduce uncertainties of the neutrino interaction
model (on which detection relies) and detector based uncertainties,
the near and far detectors should preferably use the same target
materials and technology for detection.
 
There are several aspects of the technique for long--baseline
experimentation that are generic to this whole class of experiments.
Two crucial aspects of this technique are
fine timing and precise beam aiming to a remote site
too far away to be directly measured by conventional survey.
The large distance scale drives us to rely strongly on the
Global Positioning System (GPS)~\cite{Plenary_hill_0712_ref13}
for timing synchronization.
The consequent small solid angle subtended by even a large far detector
motivate a spatially wide beam with redundant systems in place to
check its aiming stability and a well surveyed near neutrino
detector~\cite{Plenary_hill_0712_ref14}
capable of measuring the beam direction precisely.

\section{Design and overview of K2K} 
The K2K neutrino beam\cite{Plenary_hill_0712_ref15} is a horn-focused
wide band $\nu_{\mu}$ beam,
with expected spectrum peaked at about 1 GeV.
The primary beam for K2K is 12 GeV kinetic energy 
protons from the KEK proton-synchrotron \cite{Plenary_hill_0712_ref16}.
Every 2.2 s, approximately $6\times 10^{12}$ protons in nine bunches are
fast-extracted in a single turn, making a 1.1 $\rm\mu{s}$ beam spill.
These protons are focused onto a 30 mm diameter, 66 cm long
aluminum target which
is a current carrying element in the first of a pair of horn
magnets operating at 250 kA.

The near detector for K2K is actually a suite of different
detectors with varying capabilities situated in a hall 300\,m
downstream of the pion production target.
The main flux measurement is based on a one-kiloton water \Cer{}
detector (1kt) which uses the same photo-multipliers as Super--Kamiokande,
the far detector.
The design is essentially the same except for scale and the analysis
for the two detectors uses the same reconstruction algorithms.

A fine grained detector sits downstream of the 1kt. 
The first main component of it is a scintillating fiber tracker 
(SciFi)~\cite{Plenary_hill_0712_ref17} with water targets enclosed between 
layers of tracking material.
The next major component is a stack of iron plates interleaved with
drift tubes to serve as a muon range detector
(MRD)~\cite{Plenary_hill_0712_ref18}.
Both the SciFi and MRD have transverse pairs of tracking planes so
tracks can be reconstructed in three dimensions.
The SciFi has planes of plastic scintillator upstream and downstream of it
to tag incoming and outgoing particles and to record fine timing
for events in common with the MRD.

The fine grained detector as a whole is excellent for event classification
studies since it can readily distinguish quasi-elastic and inelastic 
events.
The MRD is has sufficient target mass to  provide a rate of neutrino
interactions sufficient for beam stability monitoring.
In particular, its large transverse area makes it ideal for 
beam profile monitoring.

The far detector for K2K is the Super--Kamiokande detector, probably
the best known water \Cer{} detector in the world. It is located in a mine
in Kamioka town, Gifu prefecture, Japan, 250\,km from the beamline at KEK.

\section{Results From K2K} 
Any discussion of results from K2K naturally concentrates on the 
main goal of the experiment, the search for oscillations by comparing
the near and far detector measurements. Here, we present results along
this line by first detailing the near detector measurements before 
summarizing the data reduction and analysis at the far detector.
There is not enough room here to include any deep discussion of
the other results from measurements only at the near detector, so first
we will list a few items currently being analyzed.

\begin{figure}[!hb]
\includegraphics[width=\textwidth]{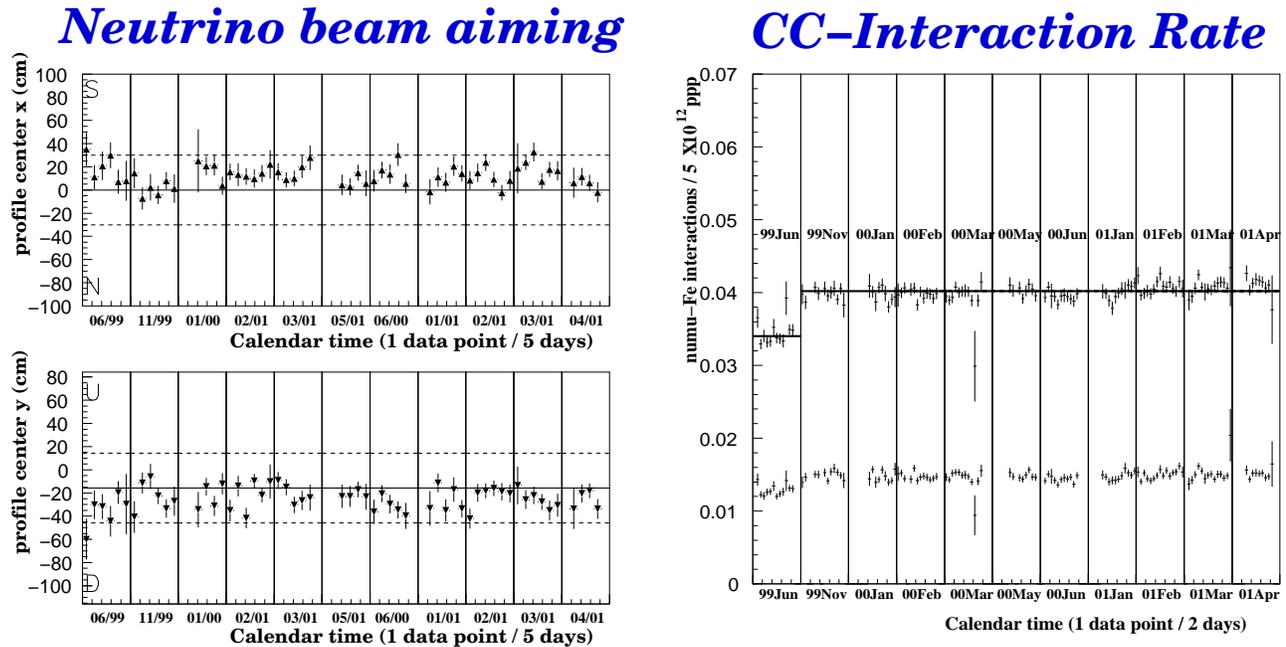}
\caption{Beam stability as measured by the muon range detector.
         Left is the centroid of reconstructed vertices each five days
         through all the data analyzed. Right is the event rate
         measured each two days. The rate in June of 1999 is expected
         to be lower than later runs since the target area 
         configuration was slightly different.
        }
\label{Plenary_hill_0712_fig2}
\end{figure}

\subsection{Topics Other than Neutrino Oscillations at K2K}
The 1kt detector can tag electromagnetic showers with good
efficiency in the momentum range of this beam. Since the beam
contamination of $\rm\nu_e$ is low ($\sim 1\%$), a sample of 
events with $\pi^0$ are straightforward to extract. Most $\pi^0$ 
events with no other visible ring are the result of neutral
current interactions ($\rm\nu N\rightarrow \nu N \pi^0$).
Therefor, this detector can be used to measure well the cross section for this 
process compared to the relatively well known quasi-elastic
cross section in this momentum range.
Such a measurement will aid atmospheric neutrino analyses
distinguishing $\rm\nu_\mu \rightarrow\nu_s$ from 
$\rm\nu_\mu \rightarrow\nu_\tau$ by the presence or absence of
neutral current events.

Events with a $\pi^0$ and a non-showering ring are likely to be
from the charged current process $\rm\nu_\mu n\rightarrow \mu p  \pi^0$,
the dominant background for an important mode of proton decay 
predicted by some super--symmetric theories\cite{Plenary_hill_0712_ref19}.
Precise measurement of this interaction cross section is crucial to
understanding the backgrounds for future even larger proton decay
experiments.

While the K2K neutrino beam consists predominantly of neutrinos from charged
pion decay, the highest energy part of the beam ($>\sim 4$GeV/c)
comes almost exclusively from kaon decays. While this accounts for
only about a percent of the neutrinos in the beam, there are enough
such events to easily tag as coming from neutrinos that must have
originated from kaons (although the geometrical acceptance is low).
Most of the nuclear physics data on meson production by proton beams
is quite old and the relative normalizations of different experiments
has a large uncertainty.
The hundred or so high energy neutrino events we already observe
can help us to limit the uncertainty on the relative meson
production in the target.

\subsection{Near Detector Results Related to Oscillation}

A prerequisite for any measurement in such an experiment is the direct
monitoring of the stability of the beam. This is especially true for
oscillation searches where aiming to the far detector is a crucial issue.
A series of charged particle monitors along the beam-line measure the
distribution of the progenitors and ``sisters'' of our neutrino beam.
Many of these give crucial information about the pulse to pulse
stability of the beam. More direct information about the neutrino beam
can be gleaned from the near detector measurements, particularly
with the large mass and area of the MRD.
Figure~\ref{Plenary_hill_0712_fig2} shows neutrino beam data from the MRD relati
ng to
the aiming and flux of neutrinos as a function of calendar time.

\begin{figure}[!ht]
\includegraphics[width=0.6\textwidth]{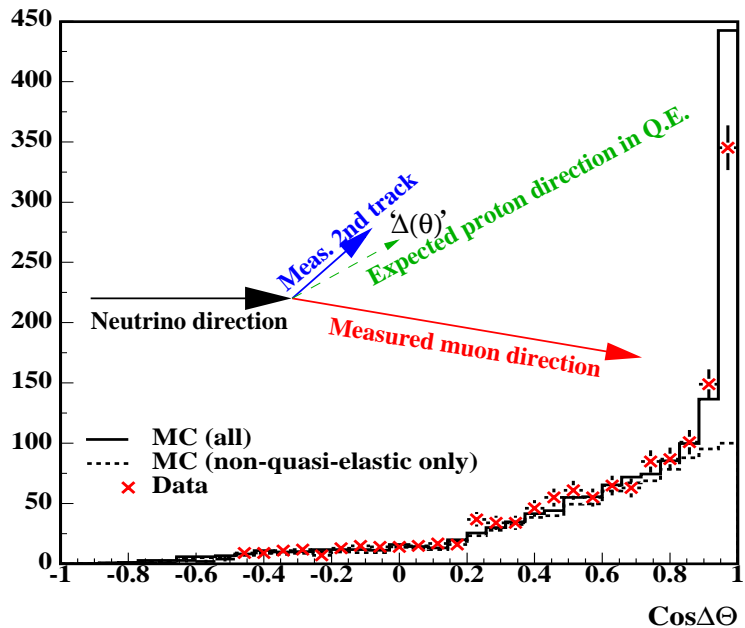}
\caption{Background measurement with 2-track events from the fine-grained
         detector. The abscissa is the cosine of the angle between
         the second track reconstructed direction and the direction calculated
         for a proton based on the assumption that the other track was a muon 
         and 2-body kinematics as is shown in the inset cartoon. Fitting
         this distribution to a pair of amplitudes times the histogram
         shapes from simulations can give a solid measurement of the
         non-quasi-elastic background in this data set.}
\label{Plenary_hill_0712_fig3}
\end{figure}

For a correct spectral measurement of the neutrino beam, background
of non-quasi-elastic interactions needs to be well understood.
The fine grained detector with scintillating fiber planes between
water targets using the MRD as a muon calorimeter is well suited 
for this study since it can reconstruct two-track kinematics with great
precision.
Figure~\ref{Plenary_hill_0712_fig3} gives some idea of how such
a measurement can be done.

\begin{figure}[!ht]
\includegraphics[width=0.9\textwidth]{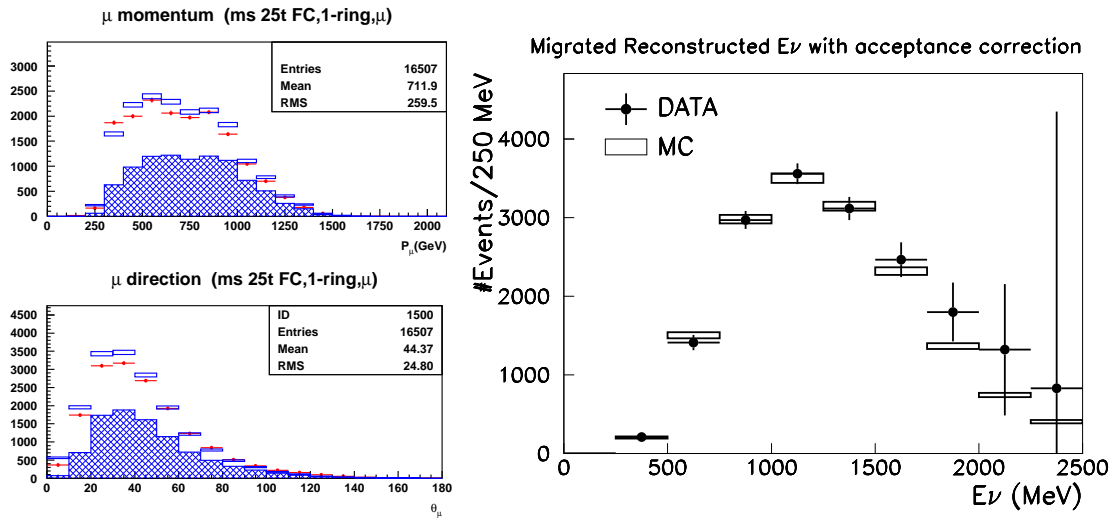}
\caption{Neutrino momentum spectrum reconstructed from data from 
         only the 1kt detector.
         Left are the basic distributions for single ring $\mu$-like
         that go into calculating a spectrum. Points are data; the
         boxes are the total prediction for simulation; the filled histogram is
         the non-quasi-elastic contribution to the simulated sample.
         Only statistical error is shown.
         Right is a neutrino spectrum after background subtraction
         and corrections for efficiencies and acceptance.
         The inclusion of information from the other detectors will
         raise the efficiency for high energy events.
         }
\label{Plenary_hill_0712_fig4}
\end{figure}

With the beam stability assured and the background measured, we can set
about measuring a spectrum at the near detectors.
The ideal detector for the spectral measurement is the 1kt detector
since it is essentially a scale model of the far detector. 
It can give the absolute flux prediction for the far detector
 with small systematic uncertainties. 
After ongoing work, particularly to understand the high energy tail
where information from different detectors must be combined, it
will provide the basis for a well-measured spectrum which can be
used in tuning our simulations.
Currently, the spectral measurement is used as a check on the
existing modeling.
One can see from figure~\ref{Plenary_hill_0712_fig4}
 that while there is some room for improvement, the inferred
spectrum is quite similar to the simulation result. Once the efficiency
for the high energy tail is better understood, we will know about
the significance of the apparent small discrepancy and tune our
simulations accordingly.

\subsection{Far Detector Results}
\begin{figure}[!ht]
\includegraphics[width=0.6\textwidth]{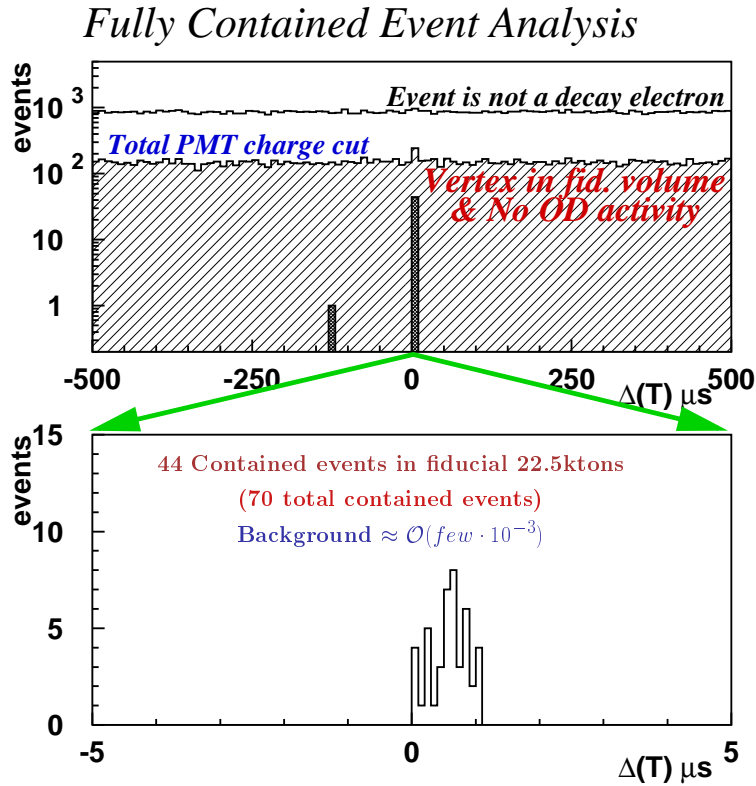}
\caption[Data reduction for SK-K2K events]{(A simplified view of)
         Data reduction for SK-K2K 
         fully-contained events.}
\label{Plenary_hill_0712_fig5}
\end{figure}

The far detector, Super--Kamiokande, aquires data constantly with 
over 90\% live time. Physics events are all self-triggered.
Timing analysis for correlation with the arrival of the K2K
beam is done off-line by comparing files of timestamps from the 
beam channel to the GPS trigger time recorded with each SK event.
The beam macro-structure of a $1.1\rm\mu s$ pulse each 2.2\,sec
gives a duty factor small enough so that an analysis for fully contained
events is virtually without background.
Figure~\ref{Plenary_hill_0712_fig5} shows a summary view of the data reduction
for fully contained event analysis.
The one millisecond window for initial selection has one event out
of time with the K2K beam. This is what is expected knowing the rate
of atmospheric neutrino detections and the sample size and duty factor.
That one event is presumably a good atmospheric neutrino event; it is
classified as a single ring showering event.
The bottom plot in the figure shows an expansion of the two central bins
of the upper plot for only the final fully contained fiducial volume
event subsample.
The definition of zero time is the expected time of arrival of the K2K
beam, so assuming a syncronization random error of less than 200\,ns,
we expect the good events to be included in a search window from
$-0.2$ to $1.3\rm\mu s$.
The factors leading to the 200\,ns error are really an upper limit,
and the fact that these data show no tail beyond the expected beam window
seems to suggest the synchronization is actually somewhat better than that. 
Table~\ref{Plenary_hill_0712_tab1} shows a breakdown of the contained,
in fiducial volume events into subclasses.

Analysis of outside fiducial volume and outer detector events gives a
total sample with which to check timing of 104 events.
The outer detector search has an expected background of order one.

{\def\Rgt{\hspace*{\stretch{1}}}
\begin{table}[!ht]
\caption[Observed FC K2K--SK events]{Observed fully-contained
         K2K--SK events with vertices in
         the fiducial volume. A total of 104 events in time with
         the expected arrival time of K2K beam includes 70 
         fully-contained events of which these are the best understood
         subset. The indentation of entries in the first two columns
         indicates that some categories are subsets of one above.
         The expectation for ``e-like'' events changes with oscillation
         since some are mis-identified $\nu_\mu$ events
         (not $\rm\nu_e$ beam contamination).}
\label{Plenary_hill_0712_tab1}
\begin{tabular}{|c|l||l|ccc|}
\hline
 Category &\  Observed\  & Exp. (no osc.) & 
                  \multicolumn{3}{c|}{Exp. with full mixing and
                      $\rm\Delta(m^2)$ in $\rm 10^{-3} eV^2 =$}\\
 & & &\makebox[21mm][c]{3.} & \makebox[21mm][c]{5.} & \makebox[21mm][c]{7.} \\
\hline
FC-22.5kton (all) & 44 & $63.9^{+6.1}_{-6.6}$ & 
              $41.5\pm 4.7$ & $27.4\pm 3.1$ & $23.1\pm 2.6$ \\
\hline
     1-ring        &\Rgt 26 \Rgt\ & $38.4\pm 5.5$  &
              $22.3\pm 3.4$ & $14.1\pm 2.2$ & $13.1\pm 2.0$ \\
\hline
\Rgt $\mu$-like &\Rgt 24 & $34.9\pm 5.5$ & 
              $19.3\pm 3.2$ & $11.6\pm 1.9$ & $10.7\pm 1.8$ \\
\cline{2-6}
\Rgt $\rm e$-like &\Rgt 2 & $3.5\pm 1.4$ & 
              $2.9\pm 1.2$ & $2.5\pm 1.0$ & $2.4\pm 1.0$ \\
\hline
multi-ring &\Rgt 18 \Rgt\ & $25.5\pm 4.3$  &
              $19.3\pm 3.4$ & $13.3\pm 2.3$ & $10.0\pm 1.8$ \\
\hline
\end{tabular}
\end{table}
}

\subsubsection{A Note About the Arrival Times of Events}
\begin{figure}
\begin{center}
\includegraphics[width=\textwidth]{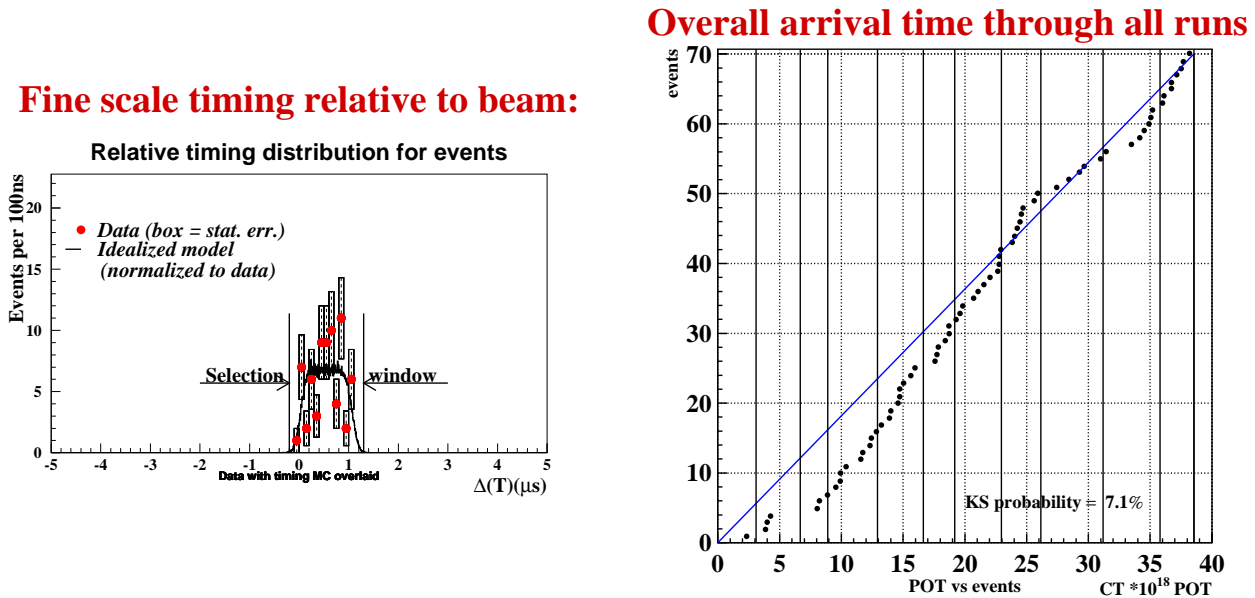}
\end{center}
\caption{(Left)The fine time distribution relative to beam arrival for
         all 70 fully contained events (points). The histogram shows
         a simple simulation of the expected distribution using the
         best knowledge of the uncertainty on synchronization.
         (Right)The time of arrival of observed far detector contained events
         in units of cumulative protons on target.
         The solid vertical lines show the boundaries of different
         run periods, all but the first of which use the same
         beamline configuration.
        }
\label{Plenary_hill_0712_fig6}
\end{figure}

A question which arises often regards the calendar time distribution
of the events observed at the far detector.
The rate of these should be proportional to the rate observed at the
near detector at any time.
To a very good approximation, this is proportional to the rate of
protons reaching the target, so the cumulative observation at the far 
detector plotted against the accumulated protons on target should 
fall roughly along a straight line.
The right plot in figure~\ref{Plenary_hill_0712_fig6} shows that distribution.
The diagonal line in 
the plot is drawn from (0,0) to the last point to show the
average rate of arrival.

The first approximately 8\% of K2K data were taken with a different
beamline configuration. This is the one point left of the first solid
vertical line. The contribution of this period to the whole result
(or even to the total observed deficit) is too small to change any
qualitative conclusion about it. The far detector expectation is
calculated using the flux information from the near detector;
the rate reduction is obvious in figure~\ref{Plenary_hill_0712_fig2}.
For reference, the overall flux measured
by the far detector is approximately 70\% of what is expected based
on near detector total flux measurements and beam simulations.

\section{Physics Interpretation of These Results} 
While a full quantative physics interpretation of these results
is not yet possible, several important statements can be made.
First, based solely on the observed number of fully contained
fiducial volume events a quantitative statement about the probability
of this observation for a given scenario involving oscillations can
be made since the error on this expectation is relatively well 
understood. Secondly, while errors on the calculation of an 
expected neutrino spectrum are not completely understood, an expected
spectrum can be constructed and from this a meaningful qualitative
statement can be made about the appearance of the observed spectral shape.

There are several ways to calculate a meaningful probability to
represent the power of a data set to make a statement about a
hypothesis. It is probably fair to say that experts agree that
the most important thing to do is to be as explicit as possible 
about how a given quantitative statement is derived.
We have tried many methods and all give consistent results.
The result of one is explained below.

A hypothesis that {\em a priori\/} predicts a central
value expected to be higher than that of competing scenarios is
disfavored at a confidence level given by: 

\[ {\rm C.L.} \equiv 1  -
    \int_{0}^{\infty}{\left({\rm e}^{-\lambda}\sum_{i=0}^{{\rm N}_{\cal O}}{
       \frac{\lambda^i}{i!}}\right) \times {\cal P}(\lambda){\rm d}\lambda}
\]
where ${\rm N}_{\cal O}$ is the number observed, and 
$\rm
 {\cal P}(\lambda) \approx \left(\sigma_L\sqrt{2\pi}\right)^{-1}
   e^{-\frac{1}{2}\left(\frac{\left(\lambda-\mu\right)}{\sigma_L}\right)^2}
$ is a gaussian approximation of the distribution of systematic
errors
for an expectation of $\mu$ with a lower side error of $\sigma_L$.

This prescription gives a probability of approximately 3\% that the
total number of fully contained fiducial volume events in the data sample
is consistent with the hypothesis of no oscillations.
This statement is independent of the information available in our
measurement of the spectrum of the observed events.
To make a quantitative statement using the spectrum, we need to 
understand the errors on the prediction of the spectrum (particularly
the correlation of errors between bins of momentum).
For now, only a qualitative statement can be made that the spectrum
reconstructed from the K2K far detector data set, while it lacks 
sufficient statistics, seems at least
as consistent with the hypothesis of oscillations with the
atmospheric neutrino best fit parameters as it does with the
no oscillations hypothesis.

\begin{wrapfigure}{c}{.45\textwidth}
\begin{center}
\includegraphics[width=0.4\textwidth]{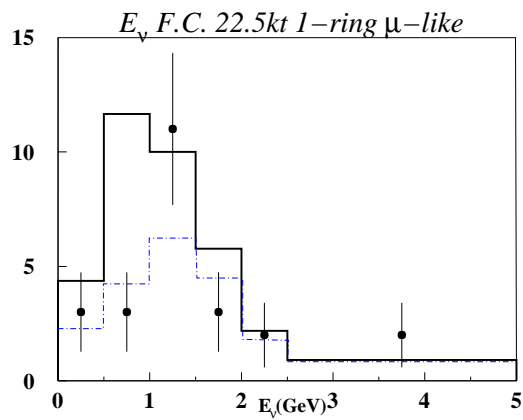}
\end{center}
\caption{The neutrino spectrum reconstructed from data (points) 
         and the prediction of simulations without (thick black solid
         line) and with (thin dot-dashed blue line) oscillations.
         The thick solid black line represents the expectations
         assuming no oscillation, and the thin dot-dashed blue line
         represents that assuming oscillations with the atmospheric
         neutrino best-fit parameters.
         }
\label{Plenary_hill_0712_fig7}
\end{wrapfigure}

Only single non-showering ring events are most likely to be produced
by quasi-elastic scatters, so only the 24 events in that subset of
fully contained events in the fiducial volume are used to make a
neutrino spectrum.
 The neutrino momentum is reconstructed from these events by assuming
quasi-elastic kinematics using the well known direction of the beam.
 Figure~\ref{Plenary_hill_0712_fig7} shows the spectrum derived from data events
along with expectations with and without oscillations. 
The expectation from our simulations are normalized by the near
detector measurement. (Later work will tune the simulation to
more closely match the near detector measured spectral shape.)
Since the systematic error on the expectation is still under study
the only errors shown are the statistical errors on the data.

\begin{figure}[!ht]
\begin{center}
\includegraphics[width=0.7\textwidth]{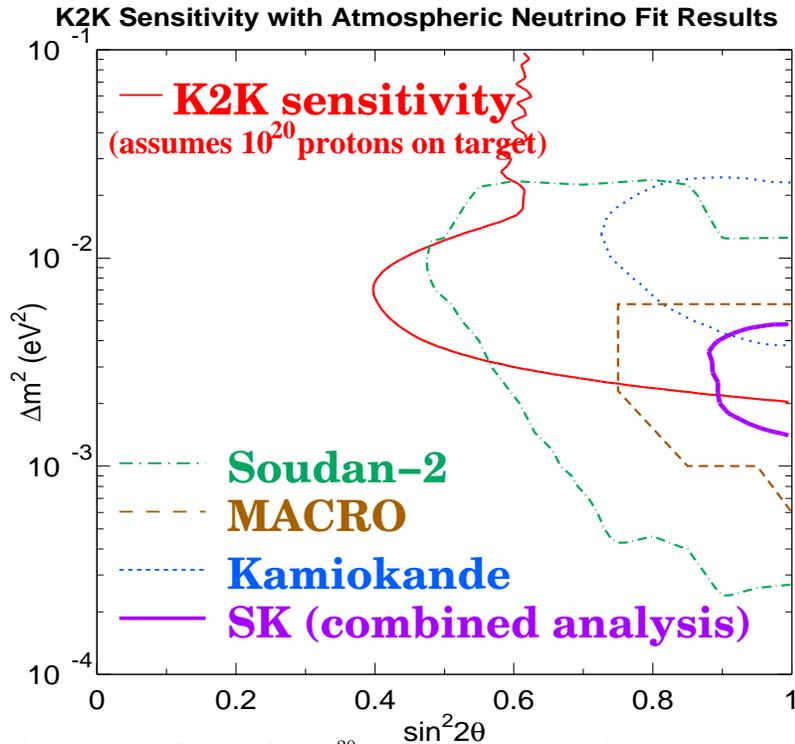}
\end{center}
\caption{The expected sensitivity of K2K after $10^{20}$ protons on 
         target and previous results of atmospheric neutrino analyses.
         The contours are to be interpreted as delimiting 90\% C.L.
         allowed regions for the atmospheric neutrino data and 90\% C.L.
         sensitive region for K2K.
         }
\label{Plenary_hill_0712_fig8} 
\end{figure}

\section{Conclusion} 
The hypothesis of no oscillations or other new physics is disfavored
at approximately the $2\sigma$ level based on analysis of the total
counts observed in K2K.
Spectral analysis gives more information, but a quantitative statement
based on it is not yet available.
Within a few years, K2K expects to gather more data and study systematic
errors to make a firm statement about oscillations.
With the full data set, K2K will cover a sensitive region of parameter
space for oscillations which overlaps significantly with the atmospheric
neutrino allowed region, giving it a good chance to  confirm the
nature of the effect.
Figure~\ref{Plenary_hill_0712_fig8} shows the (90\% C.L.) sensitive region of
parameter space for two family mixing along with the allowed regions 
from various atmospheric neutrino measurements.
K2K has analysed approximately 40\% of the data it is scheduled to
collect.

Furthermore, we should note again the significance of the 
successful operation of this experiment.
It represents the first time in high energy physics that fine time
synchronization has been successfully implemented over such large distance.
The method of timing and of stability checks for aiming established
by K2K will be used in a whole series of future experiments which
will explore the first new physics beyond the Standard Model.

\begin{acknowledgments}
K2K is a collaboration of approximately 100 physicists from Japan, 
Korea, and the U.S. hosted by both the Japanese High Energy Physics Lab,
KEK, and the Institute for Cosmic Ray Research.
We gratefully acknowledge the cooperation of the
Kamioka Mining and Smelting Company.
This work has been supported by the
Ministry of Education, Culture, Sports,
Science and Technology, Government of Japan,
the Japan Society for Promotion of Science,
the U.S. Department of Energy, the Korea
Research Foundation, and the Korea Science and Engineering Foundation.
We thank the KEK and ICRR Directorates for their strong support
and encouragement.
\end{acknowledgments}

\def\etal{{\it et al.}}

\end{document}